\documentstyle[12pt,amstex,amssymb,righttag]{article}

\def\llsymbol#1{\@llsymbol{\@nameuse{c@#1}}}
\def\@llsymbol#1{\ifcase#1\or {}\or {'}\or {''}\or {'''}\or 
   {''''}\or {'''''}\or  \else\@ctrerr\fi\relax}

\newcounter{contador}
\newcommand{\letra}{
   \stepcounter{equation}
   \setcounter{contador}{\value{equation}}
   \setcounter{equation}{0}
   \renewcommand{\theequation}{\thecontador.\alph{equation}}}
\newcommand{\antiletra}{
   \renewcommand{\theequation}{\arabic{equation}}
   \setcounter{equation}{\value{contador}}}

\def\citet{\@ifnextchar [{\@tempswatrue\@citey}{\@tempswafalse\@citey[]}}
 
\def\@citey[#1]#2{\if@filesw\immediate\write\@auxout{\string\citation{#2}}\fi
  \def\@citec{}\@cite{\@for\@cited:=#2\do
    {\@citec\def\@citec{--}\@ifundefined
       {b@\@cited}{{\bf ?}\@warning
       {Citation `\@cited' on page \thepage \space undefined}}%
\hbox{\csname b@\@cited\endcsname}}}{#1}}

\begin{document}

\thispagestyle{empty}

\vspace*{3cm}

\begin{center}
{\Large \bf G\"{o}del Type  solution with rotation, expansion and
closed time-like curves}\\[0.5cm]

by\\[0.5cm]

{\sl Carlos Pinheiro$^{\ast}$}\footnote{fcpnunes@@cce.ufes.br/maria@@gbl.com.br}\\
{\sl F.C. Khanna$^{\ast \ast\dag }$}\footnote{khanna@@phys.ualberta.ca} \\
and \\
{\sl Robert Riche$^{\ast \ast}$}

\vspace{3mm}
$^{\ast}$Universidade Federal do Esp\'{\i}rito Santo, UFES.\\
Centro de Ci\^encias Exatas\\
Av. Fernando Ferrari s/n$^{\underline{0}}$\\
Campus da Goiabeiras 29060-900 Vit\'oria ES -- Brazil.\\

$^{\ast \ast}$Theoretical Physics Institute, Dept. of Physics\\
University of Alberta,\\
Edmonton, AB T6G2J1, Canada\\
and\\
$^{\dag}$TRIUMF, 4004, Wesbrook Mall,\\
V6T2A3, Vancouver, BC, Canada.

\end{center}

\newpage
\setcounter{page}{1}
\vspace*{3cm}
\begin{center}
Abstract
\end{center}

We propose a  time-varying parameter $\underline{\alpha}$ for
G\"{o}del metric and an energy momentum tensor corresponding to
this geometry is found. To satisfy covariance arguments
time-varying gravitational and cosmological term are introduced.
The ``Einstein's equation'' for this special evolution for the Universe
are written down where expansion, rotation and closed time-like 
curves appear as a combination between standard model, G\"{o}del and steady
state properties are obtained.

\newpage
\subsection*{Introduction}

\paragraph*{}
In classical cosmology it is well-known that there are  two
different geometries for the same source as in the case  of dust 
or perfect fluid. The Friedmann-Robertson-Walker and G\"{o}del solutions
have a common source for different physics. In standard cosmology we
have an expansion of the universe but no rotation as in the G\"{o}del solution.

An open question is to know if rotation implies a violation of
causality and if closed time-like curves imply rotation.

What happens if the $\underline{\alpha}$ parameter in G\"{o}del solution
is a time dependant function? Does the universe undergo a
particular evolution to get in a special way to the present epoch?

We argue that if the $\alpha$ parameter is considered to be time dependent 
in the usual G\"{o}del Geometry it is possible to find a special
source for this new  G\"{o}del type geometry such as a ``quasi-perfect
fluid'' or a ``quasi-dust''. 

By quasi-dust it is understood that the
state equation is not characterized by the pressure $p=0$, but by
$p\approx 0$. It is assumed that if  $p\approx 0$ one can 
write an energy momentum tensor very similar to the
energy tensor for a perfect fluid.

In general if the Einstein-Hilbert lagrangean with the cosmological
term is a function of time the global covariance will be lost. But it
is assumed that the universe will get a special evolution law and if
both the cosmological term and the gravitational constant are
time-varying under these assumptions the covariance is considered
again and it is possible to find a generalization of the Einstein 
equations. A  G\"{o}del type solution could describe the universe with
expansion like in the Friedman-Robertson-Walker model plus rotation
and closed time like curves as in the case of the  G\"{o}del solution.

In accordance with \cite{um,dois,sete} the time-varying $\Lambda$ may describe the
creation of matter like the steady state model \cite{quatro}.

In reality there are two possibilities to write the time-varying
cosmological term: either one can write it in the lagrangian or one
can write it on the right side of a Einstein's equations. The second
way appears quite often \cite{sete}, but here we choose the first
possibility. Thus
we obtain a special evolution law for time varying gravitational and
cosmological term to save covariance.

An energy momentum tensor which describes the new ``G\"{o}del-type''
geometry is proposed.

It is argued that such a universe is in reality a combination
characteristics of the standard cosmology, the G\"{o}del solution
and the steady state model.

The four functions $\Lambda (t),\ k(t),\ \theta (t),\ \omega (t)$
establishing the connection among the expansion from standard cosmology
described by the scalar factor $\theta$ and the angular velocity $\omega (t)$ to
showing the rotation and the time varying $\Lambda$ and
$k$ indicating the appearence of matter inside the
universe \cite{dois,sete}, gives us the motivation to jutisfy a revived 
interest in the creation of same kind of matter in the universe. 

For large scale time we can wait to find a null
expansion, constant angular velocity, zero pressure, the usual energy
momentum tensor for dust, and the standard G\"{o}del solution, but
all the time we have a possibility of a closed time-like curve.

Nothing can be said about the
initial state of the universe as for example the  singularities for the energy
density. The conservation of energy is achieved only
asymptotically \cite{sete}.

The G\"{o}del metric is written as 
\begin{equation}
ds^2=\left(dx^0+e^{\alpha x^1}dx^2\right)^2-
(dx^1)^2-\frac{1}{2}\ e^{2\alpha x^1}(dx^2)^2-(dx^3)^2
\end{equation}

The $\alpha$ parameter is assumed to be a function of time.

The Einstein Hilbert lagrangean with cosmological term is written as
\begin{equation}
{\cal  L}=-\frac{1}{2k^2}\ R\sqrt{-g} +\Lambda (t)\sqrt{-g}\ .
\end{equation}
Here both the gravitational constant and the cosmological term 
are functions of time.

The variation principle $\displaystyle{\frac{\delta S}{\delta
g^{\mu\nu}}}$ gives us the generalized Einstein tensor
$\tilde{G}_{\mu\nu}$ as
\begin{equation}
\tilde{G}_{\mu\nu}=G_{\mu\nu}+k^2\Lambda (t)g_{\mu\nu}-\left(
\frac{2R}{k}\ \frac{\delta k}{\delta t}+2k^2\
\frac{\delta\Lambda}{\delta t}\right)\ 
\frac{\delta t}{\delta g^{\mu\nu}}
\end{equation}
where $R$ is the scalar curvature and 
$\displaystyle{\frac{\delta t}{\delta g^{\mu\nu}}}\neq 0$ for all time.

The Ricci scalar in our case is given by
\begin{equation}
R=\alpha (t)^2+2x^2\alpha{'}(t)^2+2x\alpha{''}(t)
\end{equation}
where first and second derivative of the $\alpha$ parameter  are
indicated by primes.

A particular evolution law for cosmological and gravitational term
is proposed 
\begin{equation}
\hspace*{-6cm}\Lambda (t)=\sqrt{2}\ e^{-\alpha (t)x}\Lambda_0
\end{equation}
and
\begin{equation}
k^2(t)=\frac{1}{2\sqrt{2}}\ e^{\alpha (t)x}\left(\alpha (t)^2+
2x^2\alpha '(t)^2+2x\alpha ''(t)\right)k^2_0
\end{equation}
where $\Lambda_0$ and $k_0$ are the values of cosmological and
gravitational constants at the present time.

The  equation of state for a quasi perfect fluid can be written as 
\begin{equation}
p\simeq 0\ .
\end{equation}
The fact that the pressure can be approximately zero means that the
universe could be like ``dust''.

The appropriate stress energy-momentum tensor $T_{\mu\nu}$ for an
incoherent field matter in quasi-rest is given by 
\begin{equation}
T^{\mu\nu}=\rho v^uv^{\nu}
\end{equation}
where 
\begin{equation}
v^u=\delta^u_0
\end{equation}
because we are using quasi-comoving coordinates.

The energy momentum tensor for quasi perfect fluid is given as.
\begin{equation}
\tilde{T}_{\mu\nu}=\rho\left(
\begin{array}{cccc}
1 & 0 &e^{\alpha (t)x} &0\\
&&&\\
0 & 0 & \displaystyle{\frac{1/2e^{\alpha (t)x}\left(1+2\alpha (t)x\right)\alpha{'}(t)}
{\alpha (t)^2}}& 0\\
&&&\\
e^{\alpha (t)x} & 
\displaystyle{\frac{1/2e^{\alpha (t)x}\left(1+2\alpha (t)x\right)\alpha{'}(t)}
{\alpha (t)^2}} & e^{2\alpha (t)x} & 0\\
&&&\\
0&0&0&0\\
\end{array}
\right)
\end{equation}
It is easy to verify that $\displaystyle{\lim_{t \rightarrow
\infty}}\tilde{T}_{\mu\nu}$ leads to the usual energy-momentum tensor
of perfect fluid if the $\alpha$ parameter is constant asymptotically.

The Einstein's equations are written as 
\begin{equation}
\tilde{G}_{\mu\nu}=-\frac{8\pi k(t)}{c^2}\ \tilde{T}_{\mu\nu}
\end{equation}
with the relevant components for the Einstein tensor, $G_{\mu\nu}$, given by
\begin{eqnarray}
G_{11} &=& \frac{1}{2}\left(\alpha (t)^2+2x^2\alpha{'}(t)^2+2x
\alpha{''}(t)\right)\ , \nonumber \\
G_{12} &=& \frac{1}{2}\ e^{\alpha (t)x}\left(1+2\alpha (t)x\right)
\alpha '(t)\ , \nonumber \\
G_{10} &=& \alpha (t)\alpha{'}(t)x \ , \\
G_{22} &=& \frac{3}{4}\ e^{2\alpha (t)x}\alpha (t)^2\ ,\nonumber \\
G_{20} &=& \frac{1}{2}\ e^{2\alpha (t)x}\alpha (t)^2\ , \nonumber \\
G_{33} &=& \frac{1}{2}\left(\alpha (t)^2+2x^2\alpha{'}(t)^2+2x
\alpha{''}(t)\right)\ , \nonumber \\
G_{00} &=& \frac{\alpha (t)^2}{2}\ . \nonumber 
\end{eqnarray}

Then the ``Einstein's'' equations are 
\letra
\begin{eqnarray}
&{}& \frac{2R}{k}\ \frac{\delta k}{\delta t}+2k^2\ 
\frac{\delta \Lambda}{\delta t} =0 \ , \\
&&\nonumber \\
&{}& \frac{8\pi k_0}{c^2}\ \rho =
\frac{\alpha (t)^2+\ x^2\alpha{'}(t)^2+
\ x \alpha{''}(t)}
{\sqrt{\frac{1}{2\sqrt{2}}\ 
e^{\alpha (t)x}\left(\alpha (t)^2+2x^2\alpha{'}(t)^2+
2x\alpha{''}(t)\right)}}\ , \\
&&\nonumber \\
&{}& \alpha (t)\alpha{'}(t)=0\ , \\
&&\nonumber \\
&{}& \frac{8\pi k_0}{c^2}\ \rho =\frac{\alpha
(t)^2+\frac{1}{2}x^2\alpha{'}(t)^2+\frac{1}{2}\ x
\alpha{''}(t)}
{\sqrt{\frac{1}{2\sqrt{2}}\ 
e^{\alpha (t)x}\left(\alpha (t)^2+2x^2\alpha{'}(t)^2+
2x\alpha{''}(t)\right)}}\ , \\
&&\nonumber \\
&{}& \frac{8\pi k_0}{c^2}\ \rho =\frac{\alpha
(t)^2+x^2\alpha{'}(t)^2+x
\alpha{''}(t)}
{\sqrt{\frac{1}{2\sqrt{2}}\ 
e^{\alpha (t)x}\left(\alpha (t)^2+2x^2\alpha{'}(t)^2+
2x\alpha{''}(t)\right)}}\ , \\
&&\nonumber \\
&{}& \frac{8\pi k_0}{c^2}\ \rho =\frac{\alpha(t)^2}
{\sqrt{\frac{1}{2\sqrt{2}}\ 
e^{\alpha (t)x}\left(\alpha (t)^2+2x^2\alpha{'}(t)^2+
2x\alpha{''}(t)\right)}}\ .
\end{eqnarray}
\antiletra

A strong constrant is indicated by  eq. (13.a) arising from the
special form of the energy-mometum tensor (10).

An asymptotic solution for ``Einstein's'' equation, which
describes the energy density under special conditions is found such
that 
\begin{eqnarray}
&{}&\lim_{t\rightarrow \infty}\alpha (t)\ \rightarrow  
\mbox{constant.}\ , \nonumber \\
&{}&\lim_{t\rightarrow \infty}\alpha{'}(t)\ \rightarrow 
0\ , \\
&{}&\lim_{t\rightarrow \infty}\alpha{''}(t)\ \rightarrow 
0\ .\nonumber
\end{eqnarray}

A particular solution which satisfies (14) is shown as 
\begin{equation}
\alpha (t)=\mbox{tagh}\ (t)=\frac{e^t-e^{-t}}{e^t+e^{-t}}
\end{equation}

At $t=0$ nothing can be said about the initial state of the universe.
Differently from the standard cosmology nothing is known about the
initial singularity of the universe. The initial state for
the energy density is not determined. At $t=\infty$ the energy density assumes a constant value like a
perfect fluid. 

The expansion factor for this case is
\begin{equation}
\theta =V^{\alpha}_{\boldmath{;\alpha}}=\frac{1}{\sqrt{-g}}\ 
\left(\sqrt{-g}V^{\alpha}\right)_{,\alpha}
\end{equation}
where semi-colon is a covariant derivative, coma means ordinary
derivative and $V^{\alpha}$ is the quadri-velocity given by (9).

For this case the expansion factor can be shown to be 
\begin{equation}
\theta \sim \frac{1}{2}\ \alpha{'}(t)>0
\end{equation}

Again, at $t=0$ there is no expansion of the universe, but for
$0<t<\infty$ the size of the universe is continually increasing similar
to the usual expansion in the standard cosmology. At $t=\infty$
the expansion stops again and the $\alpha$ parameter is a
constant like the usual G\"{o}del solution.

The solution (1) has rotation with angular velocity $\omega$. If we
assume that the angular velocity as in G\"{o}del solution \cite{quatro} is
given by
\begin{equation}
\omega =\sqrt{\pi k\rho}
\end{equation}
where $k$ is the gravitational constant and $\rho$ is the energy
density, the angular velocity in our case becomes.
\begin{equation}
w^2\approx \alpha (t)^2+\alpha{'}(t)^2+\alpha{''}(t)\ . 
\end{equation}
The angular velocity is zero at the beginning but it increases 
continuously for $0<t<\infty$ and finally it goes to a constant value
for $t=\infty$ exactly the same way as (18).

It may be verified that $\stackrel{\mbox{\bf .}}{\Lambda}(t)/\Lambda (t)$ is
proportional to $-\alpha{'}(t)$ which means that the cosmological term
will be decreasing in time so that it reaches a constant value. For
large scale of time there is no more expansion for the universe
but only a constant rotation.

It is important to remember that G\"{o}del solution, besides having a
constant angular velocity, has closed time-like curves. For the usual
G\"{o}del solution (1) with constant parameter, $\alpha$, the
coordinate   
system may be changed from $(x^0,x^1,x^2,x^3)$ to another system such
as $(r,\theta ,\phi ,z)$ and the metric is written as
\begin{equation}
ds^2=4a^2\left[dt^2+2H(r)d\phi dt+G(r)d\phi^2-dz^2-dr^2\right]\ , 
\end{equation}
where
\begin{eqnarray}
G(r) &=& \mbox{senh}^4(r)-\sinh^2(r)\ , \nonumber \\
\mbox{and}\ H(r) &=& 2\sqrt{2}\sinh^2(r)\ . \nonumber 
\end{eqnarray}

The G\"{o}del solution has in fact closed time like curves as we
consider $t,\ z$ and $r$ as constant \cite{quatro,seis} since $d\phi^2$ varies between
zero and $2\pi$.

Here arbitrary time dependence of functions is considered as the
usual G\"{o}del solution to be the $\underline{\alpha}$ parameter.
With an arbitrary time dependent function we get 
\begin{equation}
ds^2=4a^2dt^2+8a^2H(r)d\phi dt+4a^2G(r)m(t)d\phi^2-4a^2dz^2-4a^2dr^2\
, 
\end{equation} 
the same form as for $t,z,r$ given by constants we may find closed
time like-curves. The line element is written now as 
\begin{equation}
ds^2=4a^2G(r)m(t)d\phi^2
\end{equation}
where $a$ is a constant, and $m(t)$ some positive function. 

We can change from $(x^0,x^1,x^2,x^3)$ to a new coordinate system
$(r,\theta ,\phi ,z)$ since the covariance is guaranteed by the two
appropriate choices of cosmological term and gravitational time
dependent function.

The proposed solution carries the possibility of expansion like the
Friedmann model but it contains rotation and closed time-like curves
as in G\"{o}del solution.

The universe could get a special way of evolution in accordance with
eq. (5,6) and to continue today its course with slow expansion plus
increasing rotation and closed time like curves.

The time dependent gravitational term could mean that some kind of
matter appears in the universe exactly the same way as in the
steady state model \cite{quatro}.

We may be living in an universe which is a combination of the
standard model (expansion), steady state (appearence of matter) and
finally, G\"{o}del type solution (rotation). Asymptotically in time the
$\alpha$ parameter is constant, energy density as is usual for dust
source, zero expansion, non null rotation and a gravitational constant.

The conservation of energy will be possible only for large scale  
time. This fact is in accordance with \cite{um,tres,sete} and with our time-varying
$\Lambda$ and $k$ it becomes justifiable to revive the interest in
creation of matter.

\subsection*{Acknowledgements:}

\paragraph*{}
I would like to thank the Department of Physics, University of
Alberta for their hospitality. This work was supported by CNPq
(Governamental Brazilian Agencie for Research.

I would like to thank also Dr. Don N. Page for his kindness and attention
with  me at Univertsity of Alberta.

\end{document}